\documentclass[conference]{IEEEtran}\IEEEoverridecommandlockouts

\usepackage{amsmath,amssymb} \interdisplaylinepenalty=2500
\usepackage{cite}

\newtheorem{theorem}{Theorem}
\newtheorem{proposition}[theorem]{Proposition}

\DeclareMathOperator{\rank}{\sf rank\hspace{0.1em}}

\newcommand{\linspan}[1]{\left< #1 \right>}

\newcommand{\Fq}{\mathbb{F}_q}
\newcommand{\Fqm}{\mathbb{F}_{q^m}}

\newcommand{\mat}[1]{\begin{bmatrix} #1 \end{bmatrix}}

\newcommand{\dr}{d_{\rm R}}

\newcommand{\calA}{\mathcal{A}}

\newcommand{\calC}{\mathcal{C}}

\newcommand{\calS}{\mathcal{S}}

\newcommand{\calX}{\mathcal{X}}

\title{Universal Secure Error-Correcting Schemes for Network Coding}

\author{
\IEEEauthorblockN{Danilo Silva and Frank R. Kschischang}
\IEEEauthorblockA{Department of Electrical and Computer Engineering, University of Toronto \\
Toronto, Ontario M5S 3G4, Canada, {\{danilo, frank\}@comm.utoronto.ca}
}
}

\begin{document}
\maketitle

\begin{abstract}

This paper considers the problem of securing a linear network coding system against an adversary that is both an eavesdropper and a jammer. The network is assumed to transport $n$ packets from source to each receiver, and the adversary is allowed to eavesdrop on $\mu$ arbitrarily chosen links and also to inject up to $t$ erroneous packets into the network. The goal of the system is to achieve \emph{zero-error} communication that is information-theoretically secure from the adversary. Moreover, this goal must be attained in a universal fashion, i.e., regardless of the network topology or the underlying network code. An upper bound on the achievable rate under these requirements is shown to be $n-\mu-2t$ packets per transmission. A scheme is proposed that can achieve this maximum rate, for any $n$ and any field size $q$, provided the packet length $m$ is at least $n$ symbols. The scheme is based on rank-metric codes and admits low-complexity encoding and decoding. In addition, the scheme is shown to be optimal in the sense that the required packet length is the smallest possible among all universal schemes that achieve the maximum rate.

\end{abstract}

\section{Introduction}
\label{sec:introduction}

Consider a network implementing linear network coding for multicast \cite{Koetter.Medard2003}. The network may be subject to two types of attacks: a malicious user injects corrupt packets into the network in order to disrupt communication; an unauthorized eavesdropper intercepts packet transmissions in order to obtain as much information as possible about the transmitted messages. The linear mixing performed by network coding presents challenges to coding schemes in both scenarios, and has motivated a significant amount of research.

This paper considers the problem of dealing with the aforementioned attacks in a universal fashion, i.e., in a way that is completely independent of the network topology and the specific network code. This has the advantage of producing schemes that are compatible with noncoherent (random) network coding \cite{Ho++2006}. Also, we focus on the most stringent requirements of \emph{zero} error probability and \emph{zero} information leakage, i.e., perfectly reliable and perfectly secure (in the information-theoretic sense) communication.

Most of the previous work on this problem deals with the special cases where only error control or only security is required. A dividing assumption among these works refers to the constraints on the packet length $m$. For a system that is required to work under any packet length (in particular, under $m=1$), the error control problem has been extensively discussed in \cite{Yeung.Cai2006:Parts-1-2,Zhang2008,Yang++2008:WeightProperties} (see references therein) and the security problem has also received significant attention \cite{Cai.Yeung2002:Secure,Feldman++2004:CapacitySecure,Rouayheb.Soljanin2007}. In all of these works, the proposed solutions require knowledge of the network code, and therefore are not universal. On the other hand, universal schemes have been proposed for the case where $m$ is required to be sufficiently large; this is the approach taken in \cite{Kotter.Kschischang2008,Silva++2008} for error control and in \cite{Silva.Kschischang2008:Security-IT} for security.

When both requirements of error control and security are combined, the problem becomes harder, and a simple concatenation of an error control scheme and a security scheme may not necessarily work. The reason is that, if error control coding is followed by security coding, the overall codeword may not be robust to errors and, similarly, if security coding is followed by error control coding, the overall codeword may not be robust to eavesdropping. Previous work on this problem has been limited\footnote{except for an earlier, suboptimal version of this work. See \cite{Silva.Kschischang2008:Security-IT,Silva2009(PhD)}.} to non-universal schemes \cite{Ngai.Yang2007:DeterministicSEC,Ngai.Yeung2009:Secure-Error-Correcting}, which require knowledge of the network code.


In this paper, we propose a \emph{universal} scheme that achieves perfectly reliable and perfectly secure communication. Namely, in a network with a maxflow of $n$ packets, if at most $t$ error packets are injected in the network, and at most $\mu$ packets are observed by an eavesdropper, then our scheme can provide perfectly secure and reliable communication while achieving a rate of $k=n-2t-\mu$ packets per transmission. This rate is shown to be optimal. Note that a similar upper bound on rate has been shown \cite{Ngai.Yeung2009:Secure-Error-Correcting} in the context of non-universal network coding with $m=1$, but it does not apply to the problem considered here (since it ignores the possibility of exploiting $m>1$ in the coding scheme).

A requirement of our scheme is that the packet length $m$ must be at least $n$ symbols. We show that this value is optimal, in the sense that it is the smallest packet length of a universal scheme achieving the maximum rate.

A main tool in the design and analysis of our scheme is the theory of rank-metric codes \cite{Gabidulin1985}. We show that our scheme can benefit from existing efficient algorithms for rank-metric codes \cite{Silva++2008,Silva.Kschischang2009:FastDecoding-ISIT}, and therefore can be encoded and decoded with low complexity.

It is worth mentioning that there is another line of work that relaxes the assumption of zero error probability (requiring, instead, vanishingly small error probability) \cite{Jaggi++2008,Jaggi.Langberg2007}. In this case, even higher rates can be achieved \cite{Jaggi.Langberg2007}, however, the packet length must be asymptotically large.

The remainder of the paper is organized as follows. Section~\ref{sec:background} establishes the notation used and reviews background material on rank-metric codes and linear network coding. In Section~\ref{sec:problem-statement}, we define the problem of combined error control and security. In Section~\ref{sec:special-cases}, we review existing techniques for the special cases of either error control or security only. We also provide new results and insights for these scenarios, which will be useful for our proposed scheme. In Section~\ref{sec:proposed-scheme}, we present our scheme and show that it achieves the desired goals.
In Section~\ref{sec:converse-results}, we prove that our scheme is optimal both in the sense of maximal rate and smallest packet length. In Section~\ref{sec:noncoherent}, we discuss how the scheme can be extended to the case of noncoherent network coding. Finally, Section~\ref{sec:conclusion} presents our conclusions.

Some proofs are omitted due to lack of space. The full version of this work is being incorporated in the revised version of \cite{Silva.Kschischang2008:Security-IT}.

\section{Background}
\label{sec:background}

\subsection{Notation}
\label{ssec:notation}

Let $\Fq$ be a finite field. Let $\Fq^{n \times m}$ denote the set of all $n \times m$ matrices over $\Fq$, and set $\Fq^n = \Fq^{n \times 1}$. Let $\Fqm$ be an extension field of $\Fq$. Recall that $\Fqm$ is an $m$-dimensional vector space over $\Fq$. Thus, by fixing a basis for $\Fqm$ over $\Fq$, elements of $\Fqm$ may be viewed as (row) vectors in $\Fq^{1 \times m}$ and vice-versa. This identification will be used extensively throughout the paper. In particular, we may view a column vector in $\Fqm^n$ as a matrix in $\Fq^{n \times m}$ and vice-versa.

\subsection{Rank-Metric Codes}
\label{ssec:rank-metric-codes}

Let $X,Y \in \Fq^{n \times m}$ be matrices. The \emph{rank distance} between $X$ and $Y$ is defined as
  $\dr(X,Y) \triangleq \rank(Y - X)$.
As observed in \cite{Gabidulin1985}, the rank distance is indeed a \emph{metric}.

A \emph{rank-metric code} $\calC \subseteq \Fq^{n \times m}$ is a matrix code (i.e., a nonempty set of matrices) used in the context of the rank metric. The \emph{minimum rank distance} of $\mathcal{C}$, denoted $\dr(\mathcal{C})$, is the minimum rank distance between all pairs of distinct codewords of $\mathcal{C}$.

There is a rich coding theory for rank-metric codes that is analogous
to the classical coding theory in the Hamming metric. In particular,
the Singleton bound for the rank metric \cite{Gabidulin1985,Silva++2008} states that every rank-metric code $\mathcal{C} \subseteq \Fq^{n \times m}$ with minimum rank distance $d$ must satisfy
\begin{equation}\label{eq:singleton-bound}
|\mathcal{C}| \leq q^{\max\{n,m\} (\min\{n,m\} - d + 1)}.
\end{equation}
Codes that achieve this bound are called \emph{maximum-rank-distance} (MRD) codes and they are known to exist for all choices of parameters $q$, $n$, $m$ and $d \leq \min\{n,m\}$ \cite{Gabidulin1985}.

In the context of the bijection between $\Fq^{1 \times m}$ and $\Fqm$, a rank-metric code may described as a block code $\calC \subseteq \Fqm^n$ of length $n$ over $\Fqm$. (Note that, differently from classical coding theory, here we treat each codeword as a \emph{column} vector. However, to avoid confusion, we will keep the standard notation on generator and parity-check matrices of linear codes.)

It is particularly useful to consider \emph{linear} block codes over $\Fqm$.
For $m \geq n$, an important family of such codes was proposed by Gabidulin \cite{Gabidulin1985}. 
A \emph{Gabidulin code} is an $[n,k]$ linear code over $\Fqm$ defined
by the generator matrix
\begin{equation}\label{eq:generator-gabidulin}
G = \mat{
  g_0^{q^{0}} & g_1^{q^{0}} & \cdots & g_{n-1}^{q^{0}} \\
  g_0^{q^{1}} & g_1^{q^{1}} & \cdots & g_{n-1}^{q^{1}} \\
  \vdots & \vdots  & \ddots & \vdots \\
  g_0^{q^{k-1}} & g_1^{q^{k-1}} & \cdots & g_{n-1}^{q^{k-1}}
  }
\end{equation}
where the elements $g_0,\ldots,g_{n-1} \in \Fqm$ are linearly independent over $\Fq$. It is shown in \cite{Gabidulin1985} that the minimum rank distance of a Gabidulin code is $d = n-k+1$, so the code is MRD.

\subsection{Linear Network Coding}
\label{ssec:linear-network-coding}

The basic model for a (multicast) communication system using linear network coding is that of a finite-field matrix channel. At each channel use (generation) a source node transmits a batch of $n$ packets, each consisting of $m$ symbols from a finite field $\Fq$, which can be regarded as the rows of a matrix $X \in \Fq^{n \times m}$. Each link in the network transports a packet free of errors, and each node creates outgoing packets as $\Fq$-linear combinations of incoming packets. The specification of all such linear combinations defines the network code. The packets received by a (specific) destination node can be regarded as the rows of an $N \times m$ matrix $Y = AX$, where $A \in \Fq^{N \times n}$ is the transfer matrix that describes the linear transformations incurred by packets on route to the destination. The system is said to be \emph{coherent} if $A$ is known to each corresponding destination; otherwise, it is said to be \emph{noncoherent}.
The linear network code is said to be \emph{feasible} if every transfer matrix to a destination has rank $n$ (so that, in a coherent system, each destination is able to recover $X$).

The system described above is referred to as an
\emph{\mbox{$(n \times m,\, k)_q$} linear coded network}, where $k$ denotes the minimum rank among all transfer matrices. Thus, an $(n \times m,n)_q$ linear coded network contains a feasible network code.

\section{Problem Statement}
\label{sec:problem-statement}

For simplicity, we restrict attention to a single destination, since all the results in this paper can be immediately extended to multiple destinations. In addition, we focus on the fundamental case of coherent network coding; extensions to noncoherent network coding are described in Section~\ref{sec:noncoherent}.


The basic model for linear network coding described in Section~\ref{ssec:linear-network-coding} can be extended to incorporate packet errors. Suppose that at most $t$ errors can occur in any of the links, causing the corresponding packets to become corrupted. In this case, we will say that the network \emph{is subject to $t$ errors}.
Assuming, without loss of generality, an additive error model, the matrix received by the destination can be expressed as
\begin{equation}\nonumber\label{eq:channel-coherent}
  Y = AX + DZ
\end{equation}
where $Z \in \Fq^{t \times m}$ is a matrix consisting of the error packets injected and $D \in \Fq^{N \times t}$ is the transfer matrix from the affected links to the destination. Note that $D$ depends on the set of links in error.

This model can be further extended to include an eavesdropper adversary, in the spirit of the wiretap channel II of Ozarow and Wyner \cite{Ozarow.Wyner1985}. The eavesdropper is assumed to have access to the packets transmitted on any $\mu$ arbitrarily chosen links in the network. In this case, we will say that the network \emph{is subject to $\mu$ observations}. Let $W \in \Fq^{\mu \times m}$ be a matrix consisting of the packets observed by the eavesdropper. Then $W$ can be expressed as
\begin{equation}\nonumber
  W = BX
\end{equation}
where $B \in \Fq^{\mu \times n}$ is the transfer matrix from the source node to the eavesdropper. Note that $B$ depends on the set of intercepted links.

To ensure secure and reliable communication, the source node chooses the matrix $X$ as the (possibly stochastic) encoding of some message $S \in \calS$ (which should be recovered by the destination but not by the eavesdropper). The coding scheme is said to be \emph{zero-error} if $S$ can be uniquely determined from $Y$, i.e., $H(S|Y) = 0$. Here we assume that $A$ is a constant known to all, while $D \in \Fq^{N \times t}$ and $Z \in \Fq^{t \times m}$ are unknown random variables with unknown distributions (which may depend on $X$). A zero-error scheme, in this context, may also be called \emph{$t$-error-correcting} scheme. A scheme is said to be \emph{universally $t$-error-correcting} if it satisfies
\begin{equation}\label{eq:condition-reliability}
  H(S|Y) = 0, \quad \forall A \colon \rank A = n
\end{equation}
for any arbitrary distributions on $D$ and $Z$.
In other words, a universally $t$-error-correcting scheme must provide reliable communication for any of the choice of the (feasible) linear network code.

The coding scheme is said to be \emph{(perfectly) secret} if the eavesdropper gets no information about the message, i.e., if $I(S;W) = 0$. Note that this requirement depends on the choice of $B$. A scheme is said to be \emph{universally (perfectly) secret} under $\mu$ observations if it satisfies
\begin{equation}\label{eq:condition-security}
  I(S;W) = 0, \quad \forall B \in \Fq^{\mu \times m}.
\end{equation}
In other words, a universally secret scheme must guarantee secrecy for any choice of the linear network code.

In this paper, we are interested in schemes that are both universally $t$-error-correcting and universally secret under $\mu$ observations, i.e., schemes that satisfy both (\ref{eq:condition-reliability}) and (\ref{eq:condition-security}).

\section{Special Cases}
\label{sec:special-cases}

\subsection{Error Control Only}
\label{ssec:error-control-only}

Consider an $(n\times m,n)_q$ linear network subject to $t$ errors but $\mu=0$ observations. In this case, condition (\ref{eq:condition-security}) can be ignored.

In the case of a deterministic encoding, the following characterization is given in \cite{Silva.Kschischang2009:Metrics}.

\medskip
\begin{theorem}[{\cite{Silva.Kschischang2009:Metrics}}]\label{thm:error-correction-deterministic}
Consider a deterministic encoder mapping $S \in \calS$ to $X \in \Fq^{n \times m}$ whose image is given by $\calC \subseteq \Fq^{n \times m}$. There exists a universally $t$-error-correcting scheme with this encoder if and only if $\dr(\calC) \geq 2t+1$.
\end{theorem}
\medskip

From the Singleton bound (\ref{eq:singleton-bound}), it can be seen that the maximum rate achievable by a universally $t$-error-correcting scheme is given by
  $\max\{n,m\} (\min\{n,m\} - 2t)$
symbols per transmission, and it is achieved by an MRD code. In particular, the rate of $n-2t$ packets per transmission is achievable only if $m \geq n$.

In the case of a stochastic encoding, the result above does not necessarily hold, since it is conceivable that recovering $S$ from $Y$ does not necessarily enable the receiver to recover $X$. Still, it is possible to obtain the following equivalence result, which will be very useful in the sequel.

\medskip
\begin{theorem}\label{thm:stochastic-errors-erasures}
  Consider a stochastic encoding from $S \in \calS$ to $X \in \Fq^{n \times m}$. The encoding admits a universally $t$-error-correcting scheme if and only if it admits a zero-error scheme for the coherent channel $Y = AX$, for all full-rank $A \in \Fq^{(n-2t) \times n}$.
\end{theorem}
\begin{proof}
  Omitted due to lack of space.
\end{proof}
\medskip

Essentially, Theorem~\ref{thm:stochastic-errors-erasures} shows that any coding scheme that corrects $t$ packet errors can be modified at the decoder to instead correct $2t$ ``packet erasures'' (i.e., rank deficiency), and vice-versa.

\subsection{Security Only}
\label{ssec:security-only}

Consider an $(n \times m,n)_q$ linear coded network subject to $\mu$ observations but $t=0$ errors. In this case, $H(X|Y) = 0$; thus, condition (\ref{eq:condition-reliability}) can be replaced by $H(S|X) = 0$.

It is shown in \cite{Silva.Kschischang2008:Security-IT} that the maximum number of symbols per transmission that can be reliably communicated with a universally secret scheme is upper bounded by $m(n-\mu)$. Moreover, this rate is achievable only if $m \geq n$.

A scheme is proposed in \cite{Silva.Kschischang2008:Security-IT} that is able to achieve this maximum rate. The scheme uses Ozarow-Wyner coset coding \cite{Ozarow.Wyner1985} based on linear MRD codes. In order to describe the scheme, it is convenient to use the bijection described in Section~\ref{ssec:notation} and think of vectors in $\Fq^{1 \times m}$ as elements of the extension field $\Fqm$. Note that this is used solely to perform the encoding and decoding operations at the source and destination nodes, and has no impact in the $\Fq$-linear network coding operations performed at the internal nodes.

Let $\calC$ be an $[n,\mu]$ linear code over $\Fqm$ with parity-check matrix $H \in \Fqm^{k \times n}$, where $k = n-\mu$. Let the message be given by $S \in \Fqm^{k}$. Encoding is performed by choosing $X \in \Fqm^{n}$ uniformly at random such that $S = HX$. In other words, $S$ is viewed as a syndrome specifying a coset of $\calC$, and $X$ is chosen as a random word from that coset. Decoding is performed simply by computing $S = HX$. It is shown in \cite{Silva.Kschischang2008:Security-IT} that this scheme is universally secret if and only if $\calC$ is an MRD code and $m \geq n$.

We now describe a convenient way to perform the encoding process. Let $T \in \Fqm^{n \times n}$ be an invertible matrix such that $H$ corresponds to the first $k$ rows of $T^{-1}$. Given a message $S \in \Fqm^k$, the encoder chooses $V \in \Fqm^{(n-k)}$ uniformly at random and independently from $S$, and produces $X \in \Fqm^n$ by computing
\begin{equation}\nonumber
  X = T \mat{S \\ V}.
\end{equation}
Note that $S = HX$. It is easy to show that $H(X|S) = n-k$, i.e., $X$ is chosen uniformly at random given $S$. Thus, this encoder indeed implements a coset coding approach.

We now give a security condition based directly on the matrix $T$ rather than its inverse.

\medskip
\begin{proposition}\label{prop:security-encoder-generator-matrix}
  The encoder described above is universally secure under $\mu \leq n-k$ observations if the last $n-k$ rows of $T^T$ form a generator matrix of an $[n,n-k]$ linear MRD code over $\Fqm$ with $m \geq n$. 
\end{proposition}
\begin{proof}
  Let $G \in \Fqm^{(n-k) \times n}$ and $G_1 \in \Fqm^{k \times n}$ be such that $T^T = \mat{G_1 \\ G}$. Then
  \begin{equation}\nonumber
    \mat{I & 0 \\ 0 & I} = T^{-1} T = \mat{H \\ H_1} \mat{G_1^T & G^T} = \mat{H G_1^T & H G^T \\ H_1 G_1^T & H_1 G^T}.
  \end{equation}
  Thus, $H G^T = 0$. Since both $G$ and $H$ are full-rank, it follows that $G$ and $H$ are generator and parity-check matrices, respectively, for exactly the same code.
\end{proof}

\section{Proposed Scheme}
\label{sec:proposed-scheme}

In this section, we propose a scheme that is universally $t$-error-correcting and universally secret under $\mu$ observations. The scheme achieves a rate of $n-\mu-2t$ packets per transmission and requires the packet length $m$ to be at least $n$ symbols. The scheme can be seen as a combination of the strategies for error control and security described in Section~\ref{sec:special-cases},
designed in such a way that they can be coupled without violating conditions (\ref{eq:condition-reliability}) and (\ref{eq:condition-security}). In what follows we make use of the identification between $\Fq^{1 \times m}$ and $\Fqm$ described in Section~\ref{ssec:notation}.

Assume that $m\geq n$ and $0 < k \leq n-\mu-2t$. Let $G_0 \in \Fqm^{(k+\mu) \times n}$ be a generator matrix of an $[n,k+\mu]$ linear MRD code over $\Fqm$. Suppose that the last $\mu$ rows of $G_0$ form a generator matrix $G \in \Fq^{\mu \times n}$ of an $[n,\mu]$ linear MRD code over $\Fqm$.

Encoding proceeds as follows. Given a message $S \in \Fqm^k$, the encoder first produces an auxiliary variable
\begin{equation}\nonumber
  U = \mat{S \\ V}
\end{equation}
by choosing $V \in \Fqm^{\mu}$ is uniformly at random and independently from $S$. Then, the encoder computes
\begin{equation}\nonumber
  X = G_0^T U.
\end{equation}

Note that the mapping from $U$ to $X$ is a deterministic mapping whose image is (a subset of)
\begin{equation}\nonumber
  \calC_0 = \{G_0^T u,\, u \in \Fqm^{(k+\mu)}\}.
\end{equation}
It follows from Theorem~\ref{thm:error-correction-deterministic} that, when $X$ is transmitted over an $(n\times m,\,n)_q$ linear coded network subject to $t$ errors, the receiver can uniquely determine $U$ (and therefore $S$) if $\dr(\calC_0) > 2t$. Since $\calC_0$ is an $[n,k+\mu]$ linear MRD code over $\Fqm$, with $m \geq n$, we have that $\dr(\calC_0) = n-k-\mu+1 \geq 2t+1$. Thus, the scheme is universally $t$-error-correcting.

In particular, decoding can be performed in two steps: first, applying a decoder for $\calC_0$ in order to find $U \in \Fqm^{k + \mu}$; then, extracting the message $S$ as the first $k$ rows of $U$.

In order to prove the secrecy of the scheme, consider first an alternative interpretation. Let $T \in \Fqm^{n \times n}$ be an invertible matrix such that the last $k+\mu$ rows of $T^T$ correspond to the matrix $G_0$. Then, we have
\begin{equation}\nonumber
  X = G_0^T U = T \mat{0 \\ U} = T \mat{S' \\ V}
\end{equation}
where
\begin{equation}\nonumber
  S' = \mat{0 \\ S}.
\end{equation}
In other words, the encoder is identical to the encoder described in Section~\ref{ssec:security-only} if $S'$ is taken as the message. Furthermore, we have that the last $\mu$ rows of $T^T$ correspond to $G$, which is the generator matrix of an $[n,\mu]$ linear MRD code over $\Fqm$. Thus, by Proposition~\ref{prop:security-encoder-generator-matrix} (which holds regardless of the message distribution), we have that the scheme is universally secret under $\mu$ observations.

The above analysis proves the following result.

\medskip
\begin{theorem}
The scheme described above is universally $t$-error-correcting and universally secret under $\mu$ observations.
\end{theorem}
\medskip

Our proposed scheme relies on the assumption that a generator matrix $G_0$ for an $[n,k+\mu]$ linear MRD code $\calC_0$ exists such that its last $\mu$ rows form a generator matrix for another $[n,\mu]$ linear MRD code. It is easy to see that, if $G_0$ is taken as a generator matrix of a Gabidulin code given in the form (\ref{eq:generator-gabidulin}), then any $\mu$ consecutive rows of $G_0$ (in particular the last ones) indeed form a generator matrix of an MRD sub-code. In this case,
decoding of $\calC_0$ can be efficiently performed using the methods in \cite{Silva++2008,Silva.Kschischang2009:FastDecoding-ISIT,Silva2009(PhD)}.

%

\section{Converse Results}
\label{sec:converse-results}

In this section, we prove that our proposed scheme is optimal, both in the sense of achieving the maximum possible rate and in the sense of requiring the minimum possible packet length among all schemes that achieve this maximum rate.

\medskip
\begin{theorem}\label{thm:converse-noisy}
Consider an $(n \times m)_q$ linear coded network. Assume that the source message has entropy of $k$ packets. There exists a scheme that is universally $t$-error-correcting and universally secure under $\mu$ observations only if $k \leq n-2t-\mu$. Moreover, this maximum rate can be attained only if $m \geq n$.
\end{theorem}
\begin{proof}
Let $n' = n-2t$. Let $B \in \Fq^{\mu \times n}$ be a full-rank matrix and let $A \in \Fq^{n' \times n}$ be a full-rank matrix such that $B = PA$ for some (necessarily full-rank) $P \in \Fq^{\mu \times n'}$. Let $Y_A = AX$ and $W_B = BX = PY_A$. If the encoder admits a scheme that is universally $t$-error-correcting then, by Theorem~\ref{thm:stochastic-errors-erasures}, it also admits a scheme that is zero-error for the coherent channel $Y_A = AX$. Thus, there is a function $f_A\colon \Fq^{n' \times m} \to \calS$ such that $S = f_A(Y_A)$. In particular, there is also a function $f \colon \Fq^{n \times m} \to \calS$ such that $S = f(X)$. Thus, we may write $\calX_s = \{x \in \Fq^{n \times m}: f(x) = s\}$.
Now,
\begin{align}
k
&= H(S) \nonumber \\
&= H(S|Y_A,W_B) + I(S;Y_A,W_B) \nonumber \\
&= I(S;Y_A,W_B) \label{eq:converse-noisy-proof-1} \\
&= I(S;W_B) + I(S;Y_A|W_B) \nonumber \\
&= I(S;Y_A|W_B) \label{eq:converse-noisy-proof-2} \\
&= H(Y_A|W_B) - H(Y_A|S,W_B) \nonumber \\
&\leq H(Y_A|W_B) \label{eq:converse-noisy-proof-3} \\
&\leq n'-\rank P = n' - \mu \label{eq:converse-noisy-proof-4}
\end{align}
where (\ref{eq:converse-noisy-proof-1}) follows since $S$ is a function of $Y_A$ and (\ref{eq:converse-noisy-proof-2}) follows since $I(S;W_B) = 0$.
This proves the first statement. Now consider the second statement. Since (\ref{eq:converse-noisy-proof-4}) holds with equality, we must have $H(Y_A|S,W_B) = 0$ and $H(Y_A|W_B) = n'-\mu$. Note that these conditions hold for all full-rank $B$ and all $A \in \calA_B$, where
\begin{equation}\nonumber
  \calA_B = \{A \in \Fq^{n' \times n} : \rank A = n',\, \linspan{B} \subseteq \linspan{A}\}
\end{equation}
and $\linspan{\cdot}$ denotes the row space of a matrix.
This implies that $H(Y_A,\,A \in \calA_B | S,W_B) = 0$ and therefore $H(\bar{Y}_B | S,W_B) = 0$, where $\bar{Y}_B = \bar{A}_B X$ and $\bar{A}_B$ is the matrix consisting of the vertical stacking of all matrices in $\calA_B$. It is not hard to see that, as long as $n' > \mu$, $\rank \bar{A}_B = n$. (In fact, $\bar{A}_B$ contains every nonzero vector of $\Fq^{1 \times n}$ as one of its rows.) It follows that $H(X|S,W_B) = 0$, for all full-rank $B$. Thus, $X$ must be uniquely determined given $W_B = BX$ and the indication that $X \in \calX_S$. From Theorem~\ref{thm:error-correction-deterministic}, this implies that each $\mathcal{X}_s$ must be a rank-metric code with $\dr(\mathcal{X}_s) \geq n-\mu + 1$.

On the other hand, we have seen that $H(Y_A|W_B) = n'-\mu$ for \emph{all} full-rank $P \in \Fq^{\mu \times n'}$ where $W_B = P Y_A$ and $B=PA$. By the chain rule of entropy, it is not hard to see that this implies that $Y_A$ is uniform (for instance, by choosing some $P$'s that are submatrices of an identity matrix, as in the wiretap channel II). Thus, $H(Y_A) = n'$, which implies that $H(X) \geq n'$. Since $H(X) = H(X,S) = H(S) + H(X|S)$, we have that $H(X|S) \geq n' - k = \mu$. Thus, there must be some $s \in \calS$ such that $H(X|S=s) \geq \mu$, which implies that $|\calX_s| \geq q^{m\mu}$. Together with the fact that $\dr(\mathcal{X}_s) \geq n-\mu + 1$, we can see, from the Singleton bound (\ref{eq:singleton-bound}), that this can only happen if $m\geq n$.
\end{proof}

\section{Extension to Noncoherent Network Coding}
\label{sec:noncoherent}

The scheme described in the paper is suitable for coherent network coding and is indeed optimal. In the case of noncoherent network coding, the scheme can be adapted by including appropriate packet headers. More precisely, the transmission matrix should be $\mat{I & X}$, where $X$ is the transmission matrix of the original scheme. Clearly, including packet headers does not affect security, but it allows the scheme to be decoded when the transfer matrix $A$ is unknown. It is shown in \cite{Silva++2008} that such adaptation preserves the error-correcting capability of the code, so the universally $t$-error-correcting property is maintained. Although the rate achieved in this case is no longer optimal, it is very close to optimal for all practical packet lengths \cite{Silva++2008}.

\section{Conclusion}
\label{sec:conclusion}

In this paper, we have proposed a \emph{universal} end-to-end coding scheme that can guarantee \emph{perfectly secure} and \emph{perfectly reliable} communication over a linear coded network subject to malicious interference and eavesdropping. 
The scheme is \emph{optimal} both in the sense of achieving the maximum possible rate as well as requiring the smallest possible packet length.
The scheme is based on rank-metric codes and admit efficient encoding and decoding algorithms.


\bibliographystyle{IEEEtran}
\bibliography{IEEEabrv,networkcoding,codingtheory,rankmetric,silva,books,finitefields}

\end{document}